\documentclass{aastex}
\usepackage{spr-astr-addons}
\usepackage{url}
\usepackage{graphicx}
\usepackage{amsfonts,amssymb,amsmath}
\usepackage{esdiff}
\usepackage{color}

\RequirePackage{color}
\journal{physics Lett. B.}
\def\l{\left}
\def\r{\right}
\def\d{{\rm d}}

\def\f{\frac}
\newcommand\T{\rule{0pt}{2.6ex}}       
\newcommand\B{\rule[-1.2ex]{0pt}{0pt}} 
\def\prd{Phys. Rev. D.}
\def\jcap{JCAP}
\def\aap{Astronomy \& Astrophysics}

\def\apj{Astrophys.~J.~}
\def\apjs{Astrophys.~J.~Suppl.~Ser.~}
\def\apjl{Astrophys.~J.~Lett.~}
\def\mnras{Mon.~Not.~R.~Astron.~Soc.~}
\def\physrep{Phys. Rep.}

\def\phylB{Phy. Lett. B}
\def\phylA{Phy. Lett. A}

\def\epjc{EPJC}

\begin{document}

\title{Confronting phantom inflation with Planck data}
\slugcomment{To be submitted to Astro.Phy. \& Space Sci.}
\shorttitle{Phantom inflation with Planck data}
\shortauthors{Iqbal et al.}

\author{Asif Iqbal\altaffilmark{1}} \affil{asif@rri.res.in}  \and \author{Manzoor A. Malik\altaffilmark{2,3}}
\affil{mmalik@kashmiruniversity.ac.in}
\and
\author{Mussadiq H. Qureshi\altaffilmark{3}}
\affil{mussadiq.scholar@kashmiruniversity.net}
\altaffiltext{1}{Astronomy \& Astrophysics, Raman Research Institute, Sadashiva Nagar, Bangalore, 560080, India.}
\altaffiltext{2}{Inter-University Center for Astronomy and Astrophysics, Post Bag 4, Ganeshkhind, Pune 411007, India.}
\altaffiltext{3}{Department of Physics, University of Kashmir, Hazratbal, Srinagar, Jammu and Kashmir 190006, India.}

\begin{abstract}
The latest Planck results are in excellent agreement with the theoretical expectations predicted from standard normal inflation based on slow-roll approximation which assumes  equation-of-state $\omega \geq-1$. In this work, we study the phantom inflation ($\omega<-1$) as an alternative cosmological model within the slow-climb approximation using two hybrid inflationary fields. We perform Chain Monte Carlo analysis to determine the posterior distribution and best fit values for the cosmological parameters using Planck data and show that current CMB data does not discriminate between normal and phantom inflation. Interestingly, unlike in normal inflation, $\omega$ in phantom induced inflation  evolves very slowly away from $-1$ during the inflation. Furthermore, in contrast to the standard normal inflation for which only upper bound on tensor-to-scalar ratio $r$ are possible, we obtain both upper and lower bounds  for the two hybrid fields in the phantom scenario. Finally, we discuss prospects of future high precision polarization measurements and show that it may be possible to establish the dominance of one model over the other.
\end{abstract}


\section{Introduction}
\label{sec_1}
Precise measurements of the anisotropies in the CMB not only give us broad clue of past history of the universe but also allow us to obtain robust estimates of the cosmological parameters that govern the growth of structure of the universe. Current CMB observations are consistent with simple slow-roll model of inflation which predict adiabatic and nearly scale invariant spectrum of primordial perturbations \citep{Smoot1992,Hinshaw2013,Ade2013,Ade2015}. The  inflationary era occurred in the very early universe ($\approx 10^{-35}$ seconds after big bang) which resulted in ultra rapid stretching of the tiny patch to the observable universe. The primordial density perturbation with an almost flat spectrum (scale-invariance) were generated as direct consequence of the quantum/vacuum fluctuations produced during inflation of some scalar field \citep{Guth1981,Hawking1982,Liddle1994}. The primordial power spectrum $ \mathcal{P}(k)$  is related to the observed CMB angular power spectrum $C_{\ell}$, the position and amplitude of the peaks being highly sensitive to the cosmological parameters.

Vast number of inflationary models have been explored so far ranging from coupled scalar fields to modified gravity. Such studies are motivated by many anomalies observed in the CMB which are not consistent with the simple slow-roll inflation. These anomalies include low CMB power at large angular scales \citep{Bond1998,Iqbal2015,2017JCAP...04..013Q}, hemispheric asymmetry and the cold spot \citep{Eriksen2004,2004MNRAS.354..641H,Ade2015c}, departure from either Gaussianity or statistical isotropy \citep{2003ApJ...597L...5H,Aich2010,Ade2013b} and so on. Such models usually assume normal inflationary scenarios i.e they assume a scalar field with a positive kinetic energy term and equation  of state $\omega \geq-1$ \citep{Ade2015b} along with some additional degree of freedom.  From the last two decades or so the phantom inflationary scenario which has a negative kinetic energy term with equation-of-state $\omega <-1$ has gained some attention as an alternative cosmological model\citep{2002PhLB..545...23C,2003PhLB..562..147N,2003PhRvD..68b3522S,2004PhRvD..70f3513P,2004PhRvD..70f3521L}. Although, phantom-like forms can arise  in several theories like k-essence models \citep{1999PhLB..458..209A}, brane cosmology \citep{2003JCAP...11..014S,2008PhRvD..78b3518P}, higher order theories of gravity \citep{1988PhLB..215..635P}, string theory \citep{2014EPJC...74.3006L} such scenarios are known to suffer from the causality and stability problems \citep{2005PhRvD..72h3504B} like violating the dominated (null)  energy condition and graceful exit from the inflation. However, many approaches have been put forwad such as  quantum deSitter cosmology \citep{2003PhLB..562..147N},  gravitational back reaction \citep{2006JCAP...05..008W}, addition of additional scalar field \citep{2004PhRvD..70f3513P}   phantom-non-phantom transition \citep{2006GReGr..38.1285N,2017EPJC...77...51R}, perturbing of the isotropic and homogeneous FLRW metric and the components of the stress-energy tensor \citep{2015PhRvD..92f3019L}, effective field theories with momentum cutoff \citep{2003PhRvD..68b3509CL} etc  to avoid such problems. It has also been found that phantom models can explain the current acceleration of the universe and are consistent with classical tests of cosmology \citep{2003PhRvD..68b3522S,2004PhRvD..70d3539E,2006PhLB..632..597C,2012PhRvD..86h3008N}.

In this paper, we focus on the phantom scenario as a alternative cosmological model using two hybrid scalar fields. We obtain CMB power spectrum and corresponding primordial perturbation spectrum by carrying out MCMC analysis using the latest Planck data. We check the propensity of such models with respect to the standard normal model of inflation. The plan of this work is as follows: In Section \ref{sec_2}, we introduce phantom inflation scenario and discuss its evolution using two single scalar fields.  We discuss in section \ref{sec_3} numerical implementation used to obtain the primordial power spectrum. We then discuss methodology and CMB data set used in our analysis in section \ref{sec_4} to constrain the phantom inflation parameter space. In section \ref{sec_5}, we give parameter estimates of the inflationary models and discuss consistency of such scenarios using Planck data. Conclusion is given in section \ref{sec_6}. Throughout this work we have set $\hbar=c=8\pi G=1$ and  adopted the metric signature ($-$, $+$, $+$, $+$).

\section{Inflationary models}
\label{sec_2}
The action of the inflationary scalar field $\phi$ that is minimally coupled to gravity is given by \citep{Weinberg1972},
\begin{equation}
  S\,=\, \int d^4\,x\,\sqrt{|g|}\,\left(\frac{1}{2}R -\beta\frac{1}{2}g^{\mu \nu}\partial_{\mu}\phi \partial_{\nu}\phi + V(\phi)\right),
\end{equation}
where $\beta=-1$ for the phantom  inflation and $\beta=+1$ for normal inflation,  $R$  is the curvature scalar and $V(\phi)$ is the potential energy of the scalar field. The first term and second term in the brackets represent Lagrangians for the gravitation and scalar fields respectively.   

The energy-momentum tensor for the field $T_{\mu \nu}$ is given by \citep{Weinberg1972},
\begin{equation}
T^{\mu \nu}=\partial^\mu \phi \partial^\nu \phi-g^{\mu \nu} \mathcal{L}_\phi.
\end{equation}
Considering standard cosmological flat Friedmann-Lemaitre-Robertson-Walker (FLRW) metric \citep{1993PhR...231....1L,Liddle2000}, above equation reduces to,
\begin{eqnarray}
T^{00}&=&\rho_{\phi} = \beta\frac{1}{2} {\dot \phi}^2 + V (\phi),\\
T^{ii}&=&p_{\phi} = \beta\frac{1}{2} {\dot \phi}^2 - V (\phi),
\end{eqnarray}
where the dot stands for derivatives with respect to the cosmic time. Using equation of state (i.e $p_{\phi}=\omega \,\rho_{\phi}$) gives,
\begin{equation}
 \rho_{\phi}=\frac{\beta\,{\dot \phi}^2}{1+\omega}. 
\end{equation}
Thus, if the weak energy condition of $\rho_{\phi}\geq 0$ is to be satisfied, then for $\beta=-1$ we have phantom field having $\omega<-1$. This actually is a case where dominant energy condition, $p_{\phi}+\rho_{\phi}>0$, is violated. Moreover, considering local energy equation (i.e ${\dot \rho_\phi}=+3H(p_{\phi}+\rho_{\phi})$), we also have,
\begin{equation}
 \ddot \phi+3H\dot \phi+\beta\,V_{\phi}=0,
\label{background}
\end{equation}
where  $V_{\phi}\equiv (dV/d\phi)$. The evolution of the scalar field is coupled to the evolution of the background Friedmann equations as,
\begin{equation}
 3H^2 = \frac{1}{2} \,\beta \,{\dot \phi}^2 + V (\phi).
\end{equation}
In analogy to slow-roll normal inflation, phantom inflation is also characterized by two parameters,
\begin{equation}
 \epsilon=  -\frac{\dot H}{H^2}, \quad  \delta = - \frac{\ddot \phi}{H{\dot \phi}}.
\end{equation}
They are generally refereed to as slow-climb parameters since the phantom field is driven to up-climb in the potential unlike in normal inflation where the inflation rolls down towards  the bottom of the potentials. Moreover, unless some extra constrains are not introduced, inflationary phase driven by the phantom field, will continue up to the catastrophic Big Rip after some finite or infinite time \citep{2004MPLA...19.1509S}. However, it has been shown by many groups that graceful exit from the phantom inflation can be invoked by considering various scenarios like  additional normal scalar field \citep{2004PhRvD..70f3513P}, cosmological back reaction \citep{2006JCAP...05..008W}, decay of transient phantom field to quintessence field \citep{2017EPJC...77...51R} etc.

\begin{figure}
  \includegraphics[width=1\linewidth,height=5cm]{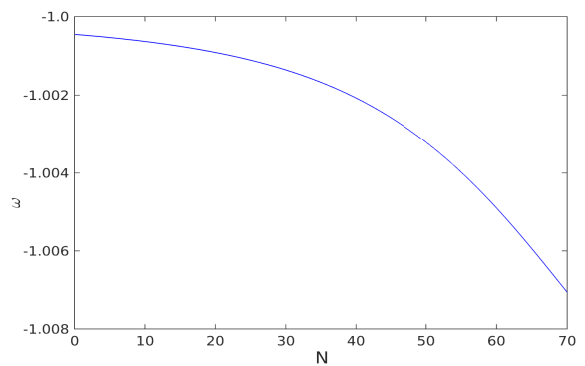}
\caption{Evolution of $\omega$ as a function of e-folds $N$ for the hybrid inflation in the phantom domain for the set of parameters which turn out to be the best fit values for the Planck data set.}
\label{omega}
\end{figure}

For the normal inflation, it has been found that small and large field inflation models always give rise to scalar spectral index of $n_s\leq1$ while as hybrid inflation can  lead to either  $n_s\geq 1$ or  $n_s\leq 1$\citep{1997PhRvD..56.3207D,2004PhRvD..70f3513P}. However, for phantom scenario, the situation is opposite and one finds that only hybrid inflation can produce spectral tilt $n_s\leq1$ (apart from $n_s>1$) \citep{2004PhRvD..70f3513P}  which is favored by the latest CMB data. Therefore, we shall consider following two hybrid inflationary potentials in the phantom domain which can produce near scale invariant primordial power spectrum,
\begin{eqnarray}
V_{\textrm{1}}(\phi)&=&V_*+\frac{1}{2}m^2\phi^2, \nonumber\\
V_{\textrm{2}}(\phi)&=&\lambda \,\left[1+\left(\frac{\phi}{\phi_*}\right)^4\right].
\label{phantom}
\end{eqnarray}
Here $V_{\textrm{1}}(\phi)$ is the quadratic potential with an additional term ($V_*$) included so as to produce the red-tilt in the scalar primordial power spectrum \citep{1997PhRvD..56.2002K,Liu2010} and $V_\textrm{2}(\phi)$ can produce either red-tilt or blue-tilt scalar power spectrum depending on the parameters used ($\lambda$, $\phi_*$). In the next section we will discuss the numerical method to obtain the primordial power  spectrum for the inflationary potentials given by Eq.~ ($\ref{phantom}$). Fig.~\ref{omega} shows the evolution of $\omega$ using best fit parameters of hybrid potential obtained in section \ref{sec_5}. Notice that during inflation the value of $\omega$ decreases further away from $-1$, however, this decline is very small, especially for the case of quadratic potential (not shown). This is in contrast with the normal inflation where strong deviations are expected in $\omega$ at the start of the inflation and its value approaches $-1$ near the end of inflation \citep{2010PhRvD..81j3502I}. Nevertheless, exact value of $\omega=-1$,  results in the scale invariance in the scaler power spectrum which is ruled out by the current CMB data at more than $5\sigma$ \citep{2014A&A...571A..22P}. 
  
We will check the performance of the above two inflationary models with respect to the standard normal inflation which assumes inflationary potential to be sufficiently flat and smooth in which case scalar and tensor spectra have simple power law form, 
\begin{eqnarray}
\ln \mathcal{P}_s(k)\, &=& \, \ln A_s\,+ (n_s-1)\ln \left(\frac{k}{k_0}\right),\\
\ln \mathcal{P}_t(k)\, &=& \, \ln A_t\,+ n_t\ln \left(\frac{k}{k_0}\right),
\end{eqnarray}
where $A_s$ ($A_t$), $n_s$  ($n_t$) are the scalar (tensor) amplitude and spectral tilt respectively and $k_0$ is the pivot scale which is set equal to $0.05$ Mpc$^{-1}$ in this work.

\section{Solving for scalar and tensor spectrum}
\label{sec_3}
The Fourier modes of the curvature perturbation ($\mathcal{R}$) and the tensor perturbation ($h$) in spatially flat universe are described as,
\begin{eqnarray}
 \mathcal{R}_k''+2\frac{z'}{z}\mathcal{R}_k'+k^2\mathcal{R}_k=0, \\
    {h}_k''+2\frac{a'}{a}{h_k}'+k^2h_k=0,
\end{eqnarray}
where  over-primes represent differentiation with respect to the conformal time and $z=a\dot \phi/H$. One usually solves the above equations along with Eq. (\ref{background}) with e-fold as the independent variable which enables us accurate and efficient numerical computation \citep{2013JCAP...05..026H}. We  assume that the initial state of inflation is built over non-phantom state with Bunch-Davies initial conditions \citep{Liu2010} on $\mathcal{R}_{k,initial}$ and $\mathcal{R'}_{k,initial}$ (or $h_{k,initial}$ and $h_{k,initial}'$), 
\begin{equation}
  (\mathcal{R}_{k,initial}, h_{k,initial})= \frac{1}{\sqrt{2\pi }z_{initial}}\exp(-i k\eta),
\end{equation}
where conformal time $\eta$ is an irrelevant phase and initial conditions are set at scales $k/aH\approx10^{2}$. Moreover, we choose the scale factor `$a_{*}$' to be such that the pivot scale $k_0=0.05$ Mpc$^{-1}$ leaves the Hubble radius at 50 e-folds before the end of the inflation \citep{2007PhRvD..76b3503H,2009PhRvD..79j3519M}. Therefore, fixing the total number of inflation e-folds to be $N_{total}=70$, as is the usual convention, implies,
\begin{equation}
 k_0=a_{*}H(N)\exp(N),
 \label{efolds}
\end{equation}
where $N=20$ and $H(N)$ is the Hubble constant at e-fold N. 

Assuming gaussianity and adiabaticity, the primordial power spectrum of curvature perturbations $\mathcal{P}_s(k)$ and tensor perturbations $\mathcal{P}_t(k)$ are given by, 
\begin{equation}
\mathcal{P}_s(k)=\frac{k^3}{2\pi^2}\left|\mathcal{R}_k\right|^2, \quad \quad \mathcal{P}_t(k)=2\frac{k^3}{2\pi^2}\left|h_k\right|^2.
\end{equation}
The additional factor of $2$ in $\mathcal{P}_t(k)$ is due to the two polarization modes of gravitational wave. The spectrum is computed at super-horizon scales (typically $k/aH\approx10^{-5}$) where it is effectively frozen. Finally, the scalar and the tensor spectral indexes are given by,
\begin{equation}
n_s=1+ \f{ \d \ln \mathcal{P}_s(k)}{\d \ln k}, \quad \quad n_t=\f{\d \ln \mathcal{P}_t(k)}{\d \ln k} .
\end{equation}
\begin{table} 
\centering
\begin{tabular}{|l|c|c|}
\hline
Parameter & Lower limit & Upper limit\\ 
\hline 
$\Omega_b{h}^2$ & 0.005 &0.1 \\ 
$\Omega_{c}{h}^2$ & 0.001 &0.99 \\ 
$\theta$ & 0.5 &10.0 \\
$\tau$ & 0.01 &0.8  \\ 
$n_s$  &  -2 & 2 \\ 
$\ln[10^{10} A_s]$ & 2.0 &4.0\\
$r$ &  0.0 &2.0 \\
${\ln}\, \l[10^{10}\, V_*\r]$ & -11.0 & -8.0\\
$\ln[10^{10}m]$ & 3.7 &5.7 \\
$\phi_*$ & 11.0 & 13.0\\
$\ln[10^{10}\lambda]$ & 0.1 &3.0 \\
\hline
\end{tabular}
\caption{Uniform prior used in parameter estimation.}
\label{priorranges}
\end{table}

\section{Observation and data analysis}
\label{sec_4}
Our analysis uses modified versions of the Boltzmann CAMB code \citep{2000ApJ...538..473L,2002PhRvD..66b3531L,1996ApJ...469..437S} and the Monte Carlo Markov Chain (MCMC) \citep{2002PhRvD..66j3511L} analysis  based CosmoMC code to calculate the theoretical CMB angular power spectra. We have used Planck 2015\footnote{Since the latest Planck results \citep{2018arXiv180706209P} (which came after the completion of this work) remarkably agree with earlier data releases we do not expect any significant changes in our results.} data set with low $\ell$ \texttt{lowl\_SMW\_70\_dx11d\_2014\_10\_03\_v5c\_Ap.clik} likelihood  ($2 \leq \ell\leq 29$) and high $\ell $ \texttt{plik\_dx11dr2\_HM\_v18\_TTTEEE.clik} likelihood ($30 \leq \ell\leq 2500$) to explore the joint likelihood ($\mathcal  L$) distribution for TT, EE, TE and BB cases. The cosmological parameterization has been carried in terms of the following parameters: baryon density ($\Omega_bh^2$), cold dark matter density ($\Omega_{c}h^2$), Thomson scattering optical depth due to re-ionization ($\tau$),  angular size of horizon ($\theta$), scalar spectral index ($n_s$), scalar amplitude ($\ln10^{10}A_s$) and tensor-to-scalar ratio at the  pivot scale of $k_0=0.05$ Mpc$^{-1}$ ($r$) along with the parameters which describe phantom inflation ($\phi_*$ and $\ln[10^{10}V_*]$ or $\ln[10^{10}m]$ and $\ln[10^{10}\lambda]$). Moreover, $n_t$ for the normal inflation under the slow roll is given by the consistency condition as $n_t=-1/8~r$. For phantom case, $n_s$ and $n_t$ is derived using Eq. 17.   The remaining parameters, which have neglible impact on the CMB power spectrum and hence cosmological parameters, were kept fixed: physical masses of standard neutrinos `$\nu$' $= 0.06$ eV, effective number of neutrinos `Neff'$=3.046$, Helium mass fraction `YHe'$=0.24$ and the width of re-ionization $=0.5$. We have used flat prior distributions and Tab.~\ref{priorranges} shows the prior ranges of all parameters used in this work.

\begin{table}
\noindent\resizebox{\linewidth}{!}{
\centering
\begin{tabular}{|c|c|c|c|c|}
\hline
Model & Parameter & Best fit & 68\% Limit&$\chi^2$\T \B\\
\hline 
Standard         & $\Omega_{\rm b}h^2$  & $0.0218$ &$0.0222^{+0.0001}_{-0.0001}$\T \B &$12944.58$ \\
 
(Normal)         & $\Omega_{\rm c}h^2$ \T \B\  &  $0.1204$ & $0.1198^{+0.0014}_{-0.0014}$ &\\

                                         & $\theta$\T \B\ & $$1.0405$$ &$1.0407^{+0.0003}_{-0.0003}$&\\

 & $\tau$\T \B\                          & $0.0717$ & $0.0771^{+0.0168}_{-0.0168}$&\\ 
  & $\ln \left[10^{10}A_s\right]$ \T \B\ &  $3.079$ & $3.0892^{+0.0327}_{-0.0327}$&\\
 & $n_s$ \T \B\                          &  $0.9649$ & $0.9649^{+0.0048}_{-0.0048}$&\\
 & $r$ \T \B\                            &  $0.0475$ & $<0.0484$&\\
\hline 
$V_1(\phi)$& $\Omega_{\rm b} h^2$        & $0.0221$ &$0.0222^{+0.0001}_{-0.0001}$\T \B &$12944.68$\\
(Phantom)& $\Omega_{\rm c}h^2$           & $0.1207$ &$0.1198^{+0.0014}_{-0.0014}$&\T \B \\
 & $\theta$                              & $1.0405$ & $1.0407^{+0.0003}_{-0.0003}$& \T \B \\
 & $\tau$                                & $0.0793$ & $0.0792^{+0.0171}_{-0.0171}$&\T \B \\ 
  & $\ln \left[10^{10}m\right]$          & $5.0304$ & $4.8561^{+0.2748}_{-0.2161}$&\T \B \\
 & $\ln \left[10^{10}V_*\right]$         & $-9.0030$ & $-9.1992^{+0.3938}_{-0.3149}$&\T \B \\

\hline
$V_2(\phi)$ & $\Omega_{\rm b}\, h^2$     & $0.0223$ & $0.0222^{+0.0001}_{-0.0001}$&$12943.94$ \T \B \\ 
(Phantom)& $\Omega_{\rm c}\, h^2$        & $0.1191$ & $0.1197^{+0.0014}_{-0.0014}$&\T \B  \\
 & $\theta$                              & $1.0410$&  $1.0407^{+0.0003}_{-0.0003}$ & \T \B \\ 
 & $\tau$                                & $0.0846$ & $0.0798^{+0.0172}_{-0.0172}$&  \T \B \\  
& ${\rm ln} \l[10^{10} \lambda\r]$       & $1.609$ &  $1.6564^{+0.4784}_{-0.4897}$ &\T \B  \\ 
 & $\phi_*$                              & $12.100$ & $12.044^{+0.3797}_{-0.1654}$&\T \B  \\
\hline 
\end{tabular}
}
\caption{The best fit  and mean values of the cosmological parameters for the two phantom driven inflationary models along with the standard normal inflation obtained using Planck data.}
\label{bestfit}
\end{table}
\section{Results and discussion}
\label{sec_5}
\begin{figure*}
\noindent\resizebox{\linewidth}{!}{
\includegraphics{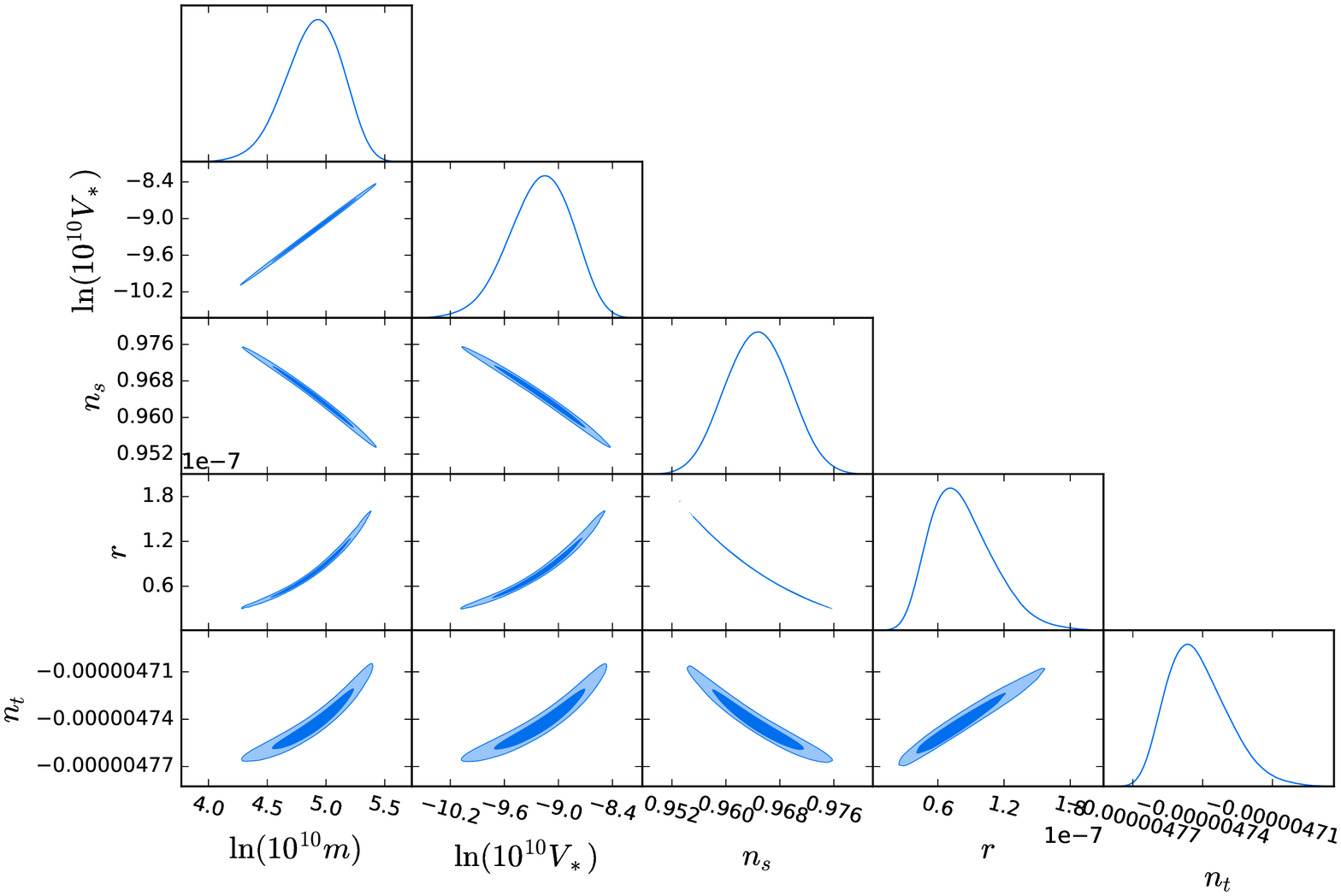}
\includegraphics{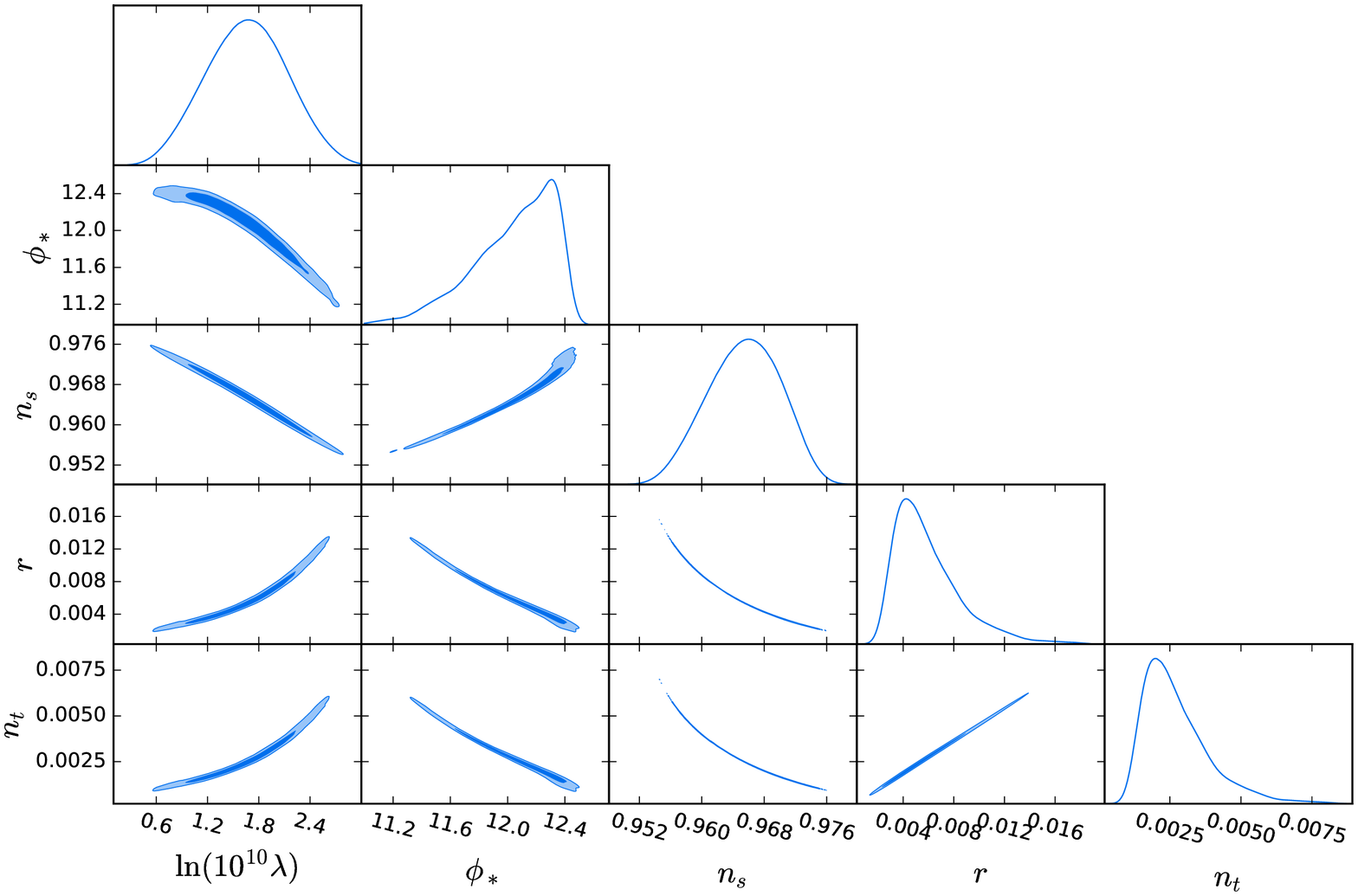}
}
\caption{Two dimensional joint posterior probability distributions and one dimensional
marginal posterior probability distribution of the inflation potential parameters and related derived parameters ($r$, $n_s$ and $n_t$).}
\label{hyb}
\end{figure*}

In this section, we present the parameter estimates of our analysis and compare the two phantom driven models against the standard normal model of inflation. Tab.~\ref{bestfit} gives the best fit values and  mean values with 1-$\sigma$ errors that we arrive at on using the inflationary parameters for the standard power law case  considering normal inflation and for the two inflationary potentials in the phantom scenario. From the chi-square estimates ($\chi^2=-2\log {\mathcal L_{max}}$, $\mathcal L_{max}$ being the maximum likelihood), which is almost same for all the three models, we find that the phantom inflation scenarios turns out to be equally favorable as standard normal inflation.
Fig.~\ref{hyb} show the marginalized posterior distribution and two dimensional posterior distributions along with contours at 68\% and 95\% of the  parameters for the two inflationary potentials in the phantom domain. It can be seen that we were able to put good constraints on the parameters for both cases. Furthermore, we also notice that the inflationary potential parameters are highly correlated. The rest of the cosmological parameters are within expected ranges and therefore are not shown in Fig.~\ref{hyb}.

\begin{figure*}
\centering
 \includegraphics[width=0.47\linewidth,height=5cm]{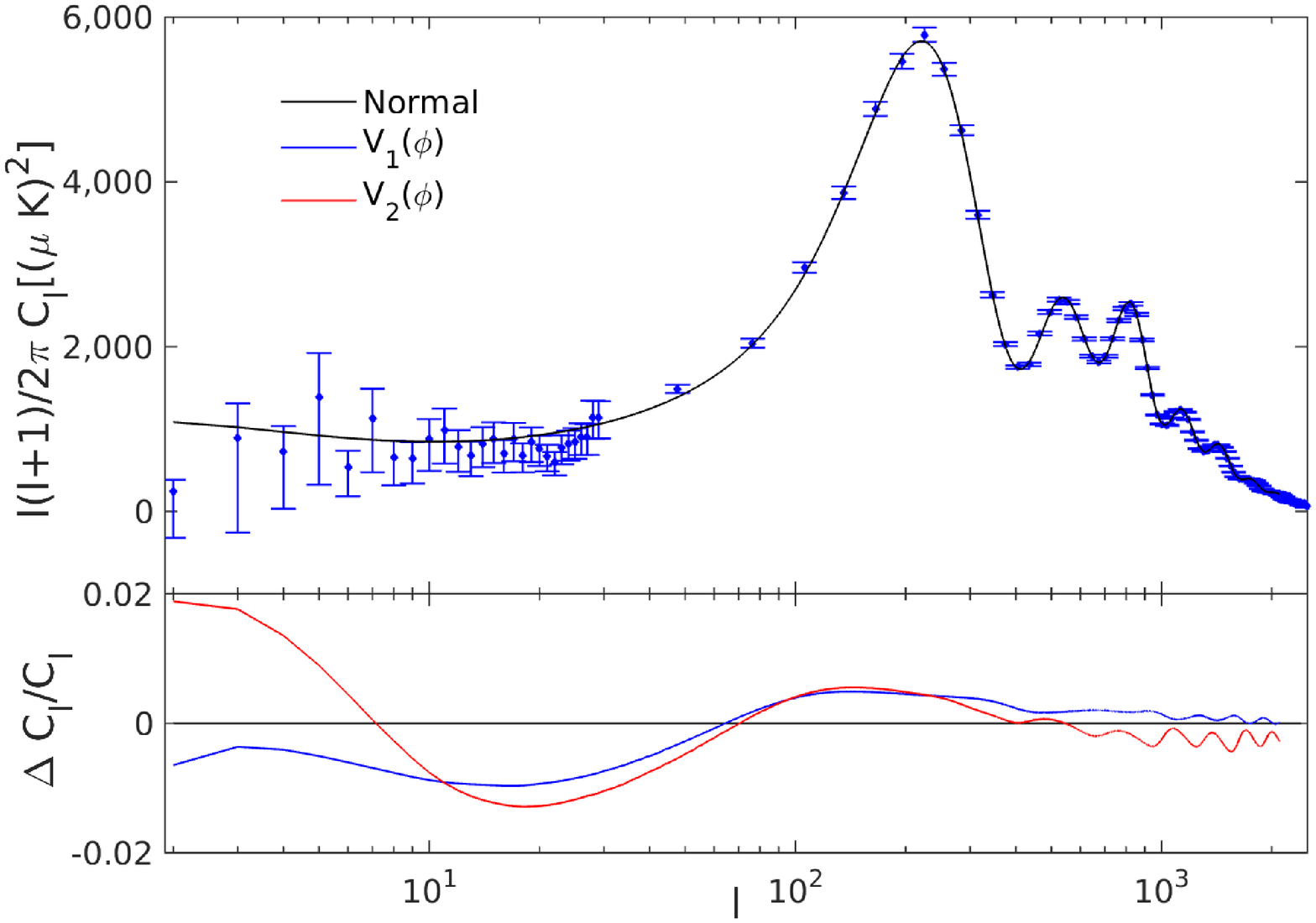}
 \includegraphics[width=0.47\linewidth,height=5cm]{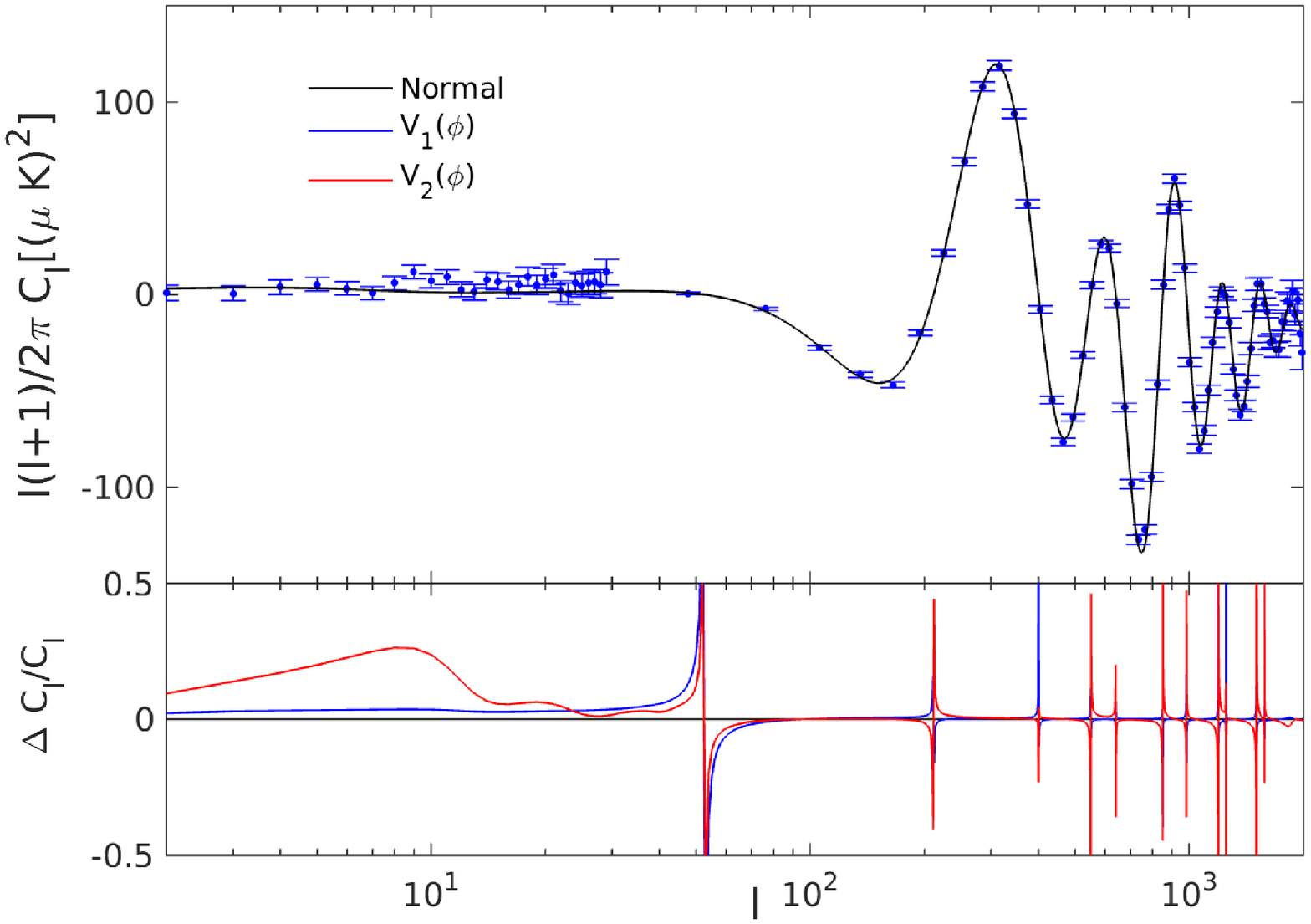}
\caption{The best CMB angular power spectrum for TT (left-panel) and TE (right-panel) modes. The observed Planck data points are shown by the blue dots. The bottom figures shows the deviations of the two inflationary models in the phantom domain from the standard model.}
\label{bestfitcl}
\end{figure*}

Fig.~\ref{bestfitcl} shows the  best fit total TT and TE CMB angular power spectra. We see that CMB power spectrum for both the inflationary potentials are strikingly similar with that implied by the normal inflation. The difference mainly occurs at the low multipoles which is cosmic variance dominated. The difference from normal inflation in TT power spectrum  is less than $2\%$  in both cases.  For TE power spectrum, we find the difference from normal inflation is at most  5\% for $V_1(\phi)$ and 20\% for $V_2(\phi)$ except at few large $\ell$ multipoles  where CMB power is close to zero (see lower panel of Fig.~\ref{bestfitcl}).

\begin{figure*}
\centering
 \includegraphics[width=0.47\linewidth,height=5cm]{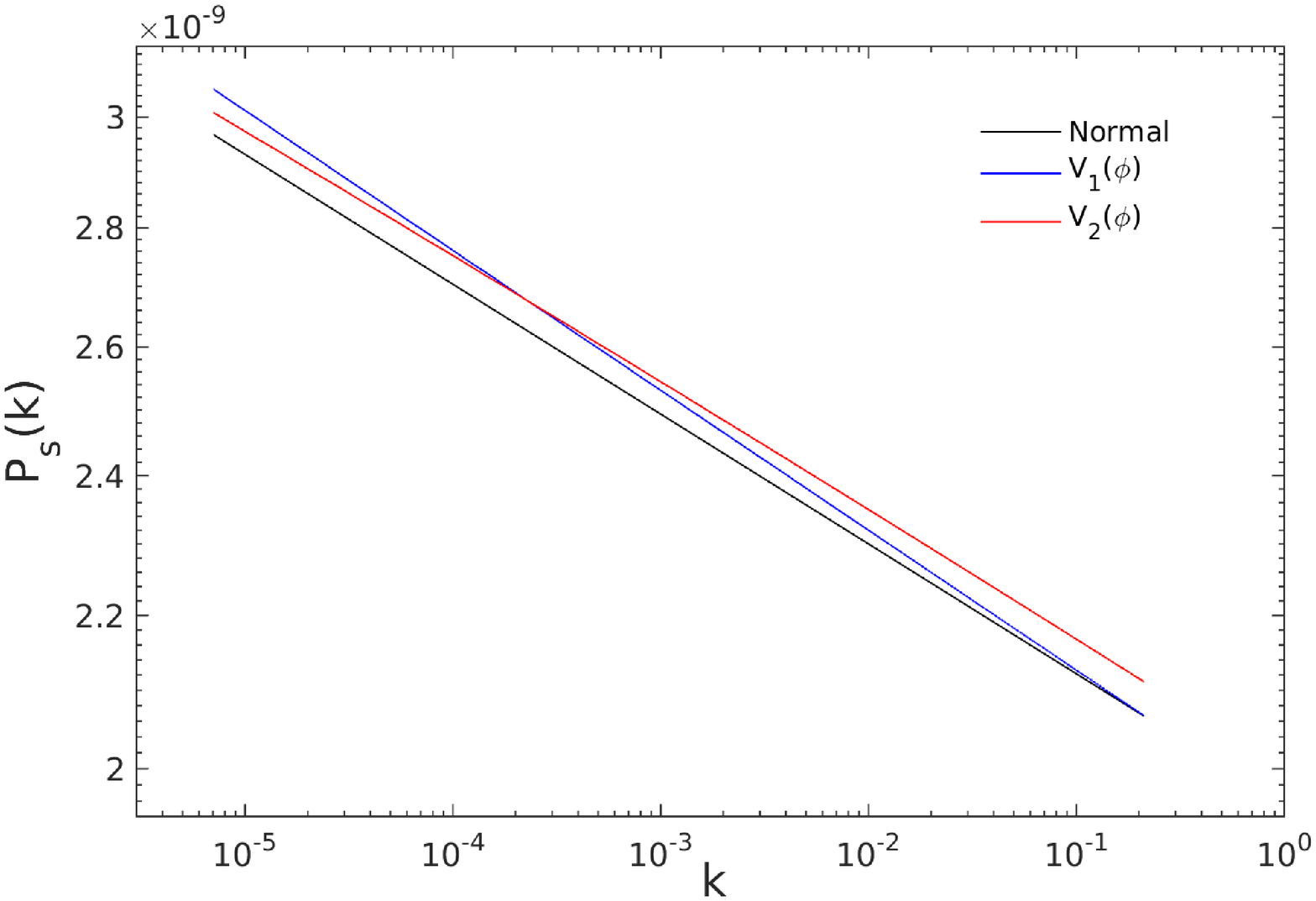}
\includegraphics[width=0.47\linewidth,height=5cm]{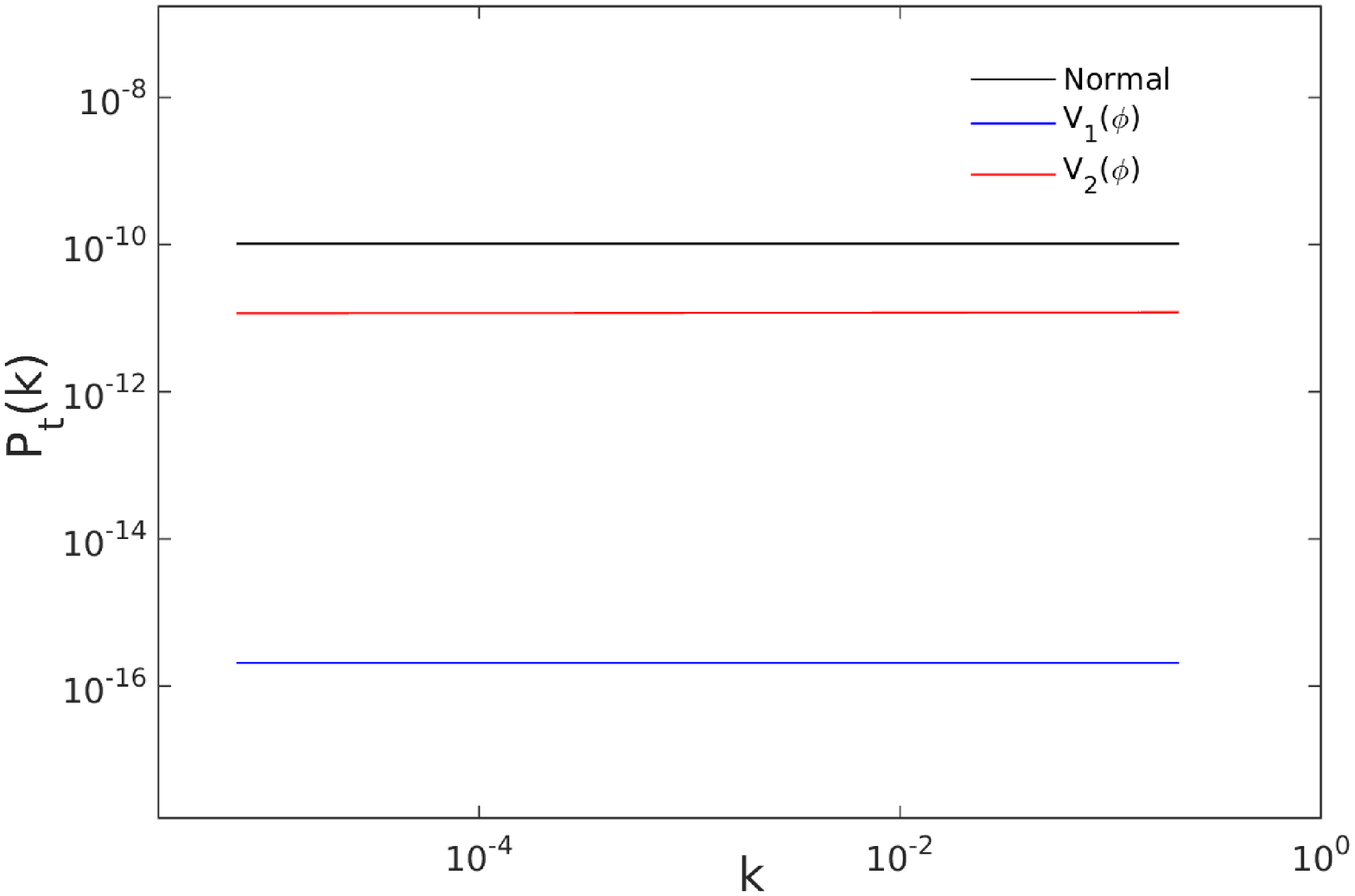}
\caption{The best fit scalar (left-panel) and tensor (right-panel) power spectrum for the CMB power spectrum shown in Figure~\ref{bestfitcl}.}
\label{bestfitpk}
\end{figure*}

However, looking at  Fig.~\ref{bestfitpk}, which shows the corresponding best fit scalar and tensor primordial spectra, one can see that tensor power spectrum for all the three cases can be easily distinguished. In particular, we see that the  best fit tensor power spectrum for the $V_1(\phi)$ is of much lower amplitude compared to other two models. For the normal inflation, we obtain only the upper limits on the tensor-to-scalar ratio $r<0.047$. On the other hand, for the phantom case, we find relatively good constrains with $r=0.823_{-0.353}^{+0.211}\times10^{-7}$ for $V_1(\phi)$ model and $r=0.591_{-0.327}^{+0.130}\times10^{-2}$ for $V_2(\phi)$ at 1-$\sigma$ confidence. It is important to note that  $r$ is better constrained in inflationary potentials because unlike in standard normal inflation, $r$ depends on the inflationary potential parameters which are constrained reasonably well. One should also mention that our inflation parameter estimates depends on the total number of inflation e-folds, $N_{total}$, chosen in Eq.~(\ref{efolds}). Since one usually presume that $60\lesssim N_{total}\lesssim80$, this dependence turns out to be small.

Moreover, interestingly phantom inflation can also result in slight blue-tilt tensor spectrum which differentiates it from the standard normal inflation \citep{2003PhRvD..68h3515P,2005PhRvD..72h3504B}. However, we find for $V_1(\phi)$ model, the tensor power spectrum has actually a red-tilt with a value of $n_t=-0.474_{-0.001}^{+0.001}\times 10^{-5}$. For $V_2(\phi)$ model, we find blue-tilt spectrum: $n_t= 0.272_{-0.145}^{+0.059}\times10^{-2}$.  It may be noted that a blue-tilted $n_t$ can also be imposed, if the propagation speed of the gravitational wave during inflation is rapidy decreasing with time \citep{2016PhRvD..93f3005C}.

Finally, although current Planck data does not distinguish the normal inflation from phantom inflation, it is worth mentioning that future CMB polarization dedicated missions such as PIXIE and S4 will have a potential to measure $r$ at the level of $r\simeq10^{-3}$  which could be decisive in distinguishing the two phantom scenarios from the standard inflation. According to the results of \citet{2017PhRvD..95f3504C}, future constraints from S4 experiment for a CMB signal with a tensor-to-scalar ratio $r \simeq 10^{-3}$ will have a sensitivity level of $\sigma_r\simeq5\times10^{-4}$ and the combination of PIXIE and S4 will enable a $5\sigma$ detection of $r$ if larger than $2\times10^{-3}$. Similar results where reported by \cite{2015JCAP...10..035H} for different future planned CMB satellite experiments.  Since, for $V_2(\phi)$ inflation we find $r\simeq10^{-3}$, future experiments should allow us to differentiate it from  normal inflation and possibly from the  quadratic inflation $V_1(\phi)$ which predict much smaller value of $r$.   Moreover, \citet{2015JCAP...10..035H} also showed that the CMB signal of tensor-to-scalar $r\simeq2\times10^{-3}$ will provide sensitivity level of $\sigma_{n_t}\simeq0.1$ on tensor spectral tilt. Since we find much smaller value for the tensor spectral tilt, therefore, it may not be possible to have any significant constraints on it so as to distinguish phantom inflation from normal inflation. 

\section{Conclusion}
\label{sec_6}
In this work, we develop a numerical routine to accurately calculate the primordial density spectrum for the physically
motivated phantom inflation.  We show that unlike in the normal inflation where $\omega$ can change rapidly towards $-1$, the phantom driven inflation actually starts with $\omega$ close to $-1$ and then varies very slowly away from $-1$ with the e-folds. We obtain the best fit estimates of CMB and primordial perturbation spectra for the two phantom driven inflation scenarios using the Planck data and compare it with the normal inflation scenario. Our results show that the evidence for the phantom-like scenarios is same as that of normal inflation. However, we find that the phantom scenarios can be differentiated from the standard normal inflation in terms of the magnitude of the tensor power spectrum  (or tensor-to-scalar ratio) which is currently not possible to precisely quantify from the present Planck data. In the light of our calculations, we show that over the next 10-15 years, a significant improvement in CMB polarization data from proposed and planned future experiments, which will have the capacity of ruling out the tensor-to-scalar ratio $r> 10^{-3}$, may provide tight constraints on the early universe particle physics models to establish a supremacy of one model over other. However, the blue-tilt tensor power spectrum of phantom inflation which separates it from normal inflation will be hard to investigate with future CMB data. 

\acknowledgments
AI and MM  would like to thank IUCAA, Pune for its hospitality. The authors would like to thank Professor Tarun Souradeep (IUCAA) for a critical scrutiny of the manuscript. We acknowledge the use of IUCAA's high performance computing facility for carrying out this work.


\begin{thebibliography}{57}
\ifx \bisbn   \undefined \def \bisbn  #1{ISBN #1}\fi
\ifx \binits  \undefined \def \binits#1{#1} \fi
\ifx \bauthor  \undefined \def \bauthor#1{#1} \fi
\ifx \batitle  \undefined \def \batitle#1{#1} \fi
\ifx \bjtitle  \undefined \def \bjtitle#1{#1}\fi
\ifx \bvolume  \undefined \def \bvolume#1{\textbf{#1}}\fi
\ifx \byear  \undefined \def \byear#1{#1} \fi
\ifx \bissue  \undefined \def \bissue#1{#1} \fi
\ifx \bfpage  \undefined \def \bfpage#1{#1} \fi
\ifx \blpage  \undefined \def \blpage #1{#1} \fi
\ifx \burl  \undefined \def \burl#1{\textsf{#1}} \fi
\ifx \doiurl  \undefined \def \doiurl#1{\textsf{#1}} \fi
\ifx \betal  \undefined \def \betal{\textit{et al.}} \fi
\ifx \binstitute  \undefined \def \binstitute#1{#1} \fi
\ifx \binstitutionaled  \undefined \def \binstitutionaled#1{#1} \fi
\ifx \bctitle  \undefined \def \bctitle#1{#1} \fi
\ifx \beditor  \undefined \def \beditor#1{#1} \fi
\ifx \bpublisher  \undefined \def \bpublisher#1{#1} \fi
\ifx \bbtitle  \undefined \def \bbtitle#1{#1} \fi
\ifx \bedition  \undefined \def \bedition#1{#1} \fi
\ifx \bseriesno  \undefined \def \bseriesno#1{#1} \fi
\ifx \blocation  \undefined \def \blocation#1{#1} \fi
\ifx \bsertitle  \undefined \def \bsertitle#1{#1} \fi
\ifx \bsnm \undefined \def \bsnm#1{#1} \fi
\ifx \bsuffix \undefined \def \bsuffix#1{#1} \fi
\ifx \bparticle \undefined \def \bparticle#1{#1} \fi
\ifx \barticle \undefined \def \barticle#1{#1} \fi
\ifx \bconfdate \undefined \def \bconfdate #1{#1} \fi
\ifx \botherref \undefined \def \botherref #1{#1} \fi
\ifx \url \undefined \def \url#1{\textsf{#1}} \fi
\ifx \bchapter \undefined \def \bchapter#1{#1} \fi
\ifx \bbook \undefined \def \bbook#1{#1} \fi
\ifx \bcomment \undefined \def \bcomment#1{#1} \fi
\ifx \oauthor \undefined \def \oauthor#1{#1} \fi
\ifx \citeauthoryear \undefined \def \citeauthoryear#1{#1} \fi
\ifx \endbibitem  \undefined \def \endbibitem {}\fi
\ifx \bconflocation  \undefined \def \bconflocation#1{#1} \fi
\ifx \arxivurl  \undefined \def \arxivurl#1{\textsf{#1}} \fi

\bibitem[\protect\citeauthoryear{{Aich} and {Souradeep}}{2010}]{Aich2010}
\begin{barticle}
\bauthor{\bsnm{{Aich}}, \binits{M.}},
\bauthor{\bsnm{{Souradeep}}, \binits{T.}}:
\bjtitle{\prd}
\bvolume{81},
\bfpage{083008}
(\byear{2010}).
\arxivurl{arXiv:1001.1723}.
doi:\doiurl{10.1103/PhysRevD.81.083008}
\end{barticle}
\endbibitem

\bibitem[\protect\citeauthoryear{{Armend\'ariz-Pic\'on}
  et~al.}{1999}]{1999PhLB..458..209A}
\begin{barticle}
\bauthor{\bsnm{{Armend\'ariz-Pic\'on}}, \binits{C.}},
\bauthor{\bsnm{{Damour}}, \binits{T.}},
\bauthor{\bsnm{{Mukhanov}}, \binits{V.}}:
\bjtitle{\phylB}
\bvolume{458},
\bfpage{209}
(\byear{1999}).
\arxivurl{arXiv:hep-th/9904075}.
doi:\doiurl{10.1016/S0370-2693(99)00603-6}
\end{barticle}
\endbibitem

\bibitem[\protect\citeauthoryear{{Baldi} et~al.}{2005}]{2005PhRvD..72h3504B}
\begin{barticle}
\bauthor{\bsnm{{Baldi}}, \binits{M.}},
\bauthor{\bsnm{{Finelli}}, \binits{F.}},
\bauthor{\bsnm{{Matarrese}}, \binits{S.}}:
\bjtitle{\prd}
\bvolume{72},
\bfpage{083504}
(\byear{2005}).
\arxivurl{arXiv:0505552}.
doi:\doiurl{10.1103/PhysRevD.72.083504}
\end{barticle}
\endbibitem

\bibitem[\protect\citeauthoryear{{Bond} et~al.}{1998}]{Bond1998}
\begin{barticle}
\bauthor{\bsnm{{Bond}}, \binits{J. R.}},
\bauthor{\bsnm{{Jaffe}}, \binits{A. H.}},
\bauthor{\bsnm{{Knox}}, \binits{L.}}:
\bjtitle{\prd}
\bvolume{57},
\bfpage{2117}
(\byear{1998}).
\arxivurl{arXiv:9708203}.
doi:\doiurl{10.1103/PhysRevD.57.2117}
\end{barticle}
\endbibitem

\bibitem[\protect\citeauthoryear{{Cai} et~al.}{2016}]{2016PhRvD..93f3005C}
\begin{barticle}
\bauthor{\bsnm{{Cai}}, \binits{Y.}},
\bauthor{\bsnm{{Wang}}, \binits{Y.-T.}},
\bauthor{\bsnm{{Piao}}, \binits{Y.-S.}}:
\bjtitle{\prd}
\bvolume{93},
\bfpage{063005}
(\byear{2016}).
\arxivurl{arXiv:1510.08716}.
doi:\doiurl{10.1103/PhysRevD.93.063005}
\end{barticle}
\endbibitem

\bibitem[\protect\citeauthoryear{{Calabrese}
  et~al.}{2017}]{2017PhRvD..95f3504C}
\begin{barticle}
\bauthor{\bsnm{{Calabrese}}, \binits{E.}},
\bauthor{\bsnm{{Alonso}}, \binits{D.}},
\bauthor{\bsnm{{Dunkley}}, \binits{J.}}:
\bjtitle{\prd}
\bvolume{95},
\bfpage{063504}
(\byear{2017}).
\arxivurl{arXiv:1611.10269}.
doi:\doiurl{10.1103/PhysRevD.95.063504}
\end{barticle}
\endbibitem

\bibitem[\protect\citeauthoryear{{Caldwell}}{2002}]{2002PhLB..545...23C}
\begin{barticle}
\bauthor{\bsnm{{Caldwell}}, \binits{R. R.}}:
\bjtitle{\phylB}
\bvolume{545},
\bfpage{23}
(\byear{2002}).
\arxivurl{arXiv:9908168}.
doi:\doiurl{10.1016/S0370-2693(02)02589-3}
\end{barticle}
\endbibitem

\bibitem[\protect\citeauthoryear{{Capozziello}
  et~al.}{2006}]{2006PhLB..632..597C}
\begin{barticle}
\bauthor{\bsnm{{Capozziello}}, \binits{S..}},
\bauthor{\bsnm{{Nojiri}}, \binits{S.}},
\bauthor{\bsnm{{Odintsov}}, \binits{S. D.}}:
\bjtitle{\phylB}
\bvolume{632},
\bfpage{597}
(\byear{2006}).
\arxivurl{arXiv:0507182}.
doi:\doiurl{10.1016/j.physletb.2005.11.012}
\end{barticle}
\endbibitem

\bibitem[\protect\citeauthoryear{{Carroll} et~al.}{2003}]{2003PhRvD..68b3509CL}
\begin{barticle}
\bauthor{\bsnm{{Carroll}}, \binits{S. M.}},
\bauthor{\bsnm{{Hoffman}}, \binits{M.}},
\bauthor{\bsnm{{Trodden}}, \binits{M.}}:
\bjtitle{\prd}
\bvolume{68},
\bfpage{023509}
(\byear{2003}).
\arxivurl{arXiv:03012732}.
doi:\doiurl{10.1103/PhysRevD.68.023509}
\end{barticle}
\endbibitem

\bibitem[\protect\citeauthoryear{{Dodelson} et~al.}{1997}]{1997PhRvD..56.3207D}
\begin{barticle}
\bauthor{\bsnm{{Dodelson}}, \binits{S.}},
\bauthor{\bsnm{{Kinney}}, \binits{W. H.}},
\bauthor{\bsnm{W.}, \binits{K. E.}}:
\bjtitle{\prd}
\bvolume{56},
\bfpage{3207}
(\byear{1997}).
\arxivurl{arXiv:9702166}.
doi:\doiurl{10.1103/PhysRevD.56.3207}
\end{barticle}
\endbibitem

\bibitem[\protect\citeauthoryear{{Elizalde} et~al.}{2004}]{2004PhRvD..70d3539E}
\begin{barticle}
\bauthor{\bsnm{{Elizalde}}, \binits{E.}},
\bauthor{\bsnm{{Nojiri}}, \binits{S.}},
\bauthor{\bsnm{{Odintsov}}, \binits{S. D.}}:
\bjtitle{\prd}
\bvolume{70},
\bfpage{043539}
(\byear{2004}).
\arxivurl{arXiv:0405034}.
doi:\doiurl{10.1103/PhysRevD.70.043539}
\end{barticle}
\endbibitem

\bibitem[\protect\citeauthoryear{{Eriksen} et~al.}{2004}]{Eriksen2004}
\begin{barticle}
\bauthor{\bsnm{{Eriksen}}, \binits{H. K.}},
\bauthor{\bsnm{{Hansen}}, \binits{F. K.}},
\bauthor{\bsnm{{Banday}}, \binits{A. J.}},
\bauthor{\bsnm{{Gorski}}, \binits{K. M.}},
\bauthor{\bsnm{{Lilje}}, \binits{P. B.}}:
\bjtitle{\apj}
\bvolume{605},
\bfpage{14}
(\byear{2004}).
\arxivurl{arXiv:0307507}.
doi:\doiurl{10.1086/382267}
\end{barticle}
\endbibitem

\bibitem[\protect\citeauthoryear{{Guth}}{1981}]{Guth1981}
\begin{barticle}
\bauthor{\bsnm{{Guth}}, \binits{A. H.}}:
\bjtitle{\prd}
\bvolume{23},
\bfpage{347}
(\byear{1981}).
doi:\doiurl{10.1103/PhysRevD.23.347}
\end{barticle}
\endbibitem

\bibitem[\protect\citeauthoryear{{Hajian} and
  {Souradeep}}{2003}]{2003ApJ...597L...5H}
\begin{barticle}
\bauthor{\bsnm{{Hajian}}, \binits{A.}},
\bauthor{\bsnm{{Souradeep}}, \binits{T.}}:
\bjtitle{\apjl}
\bvolume{597},
\bfpage{5}
(\byear{2003}).
\arxivurl{arXiv:0308001}.
doi:\doiurl{10.1086/379757}
\end{barticle}
\endbibitem

\bibitem[\protect\citeauthoryear{{Hamann} et~al.}{2007}]{2007PhRvD..76b3503H}
\begin{barticle}
\bauthor{\bsnm{{Hamann}}, \binits{J.}},
\bauthor{\bsnm{{Covi}}, \binits{L.}},
\bauthor{\bsnm{{Melchiorri}}, \binits{A.}},
\bauthor{\bsnm{{Slosar}}, \binits{A.}}:
\bjtitle{\prd}
\bvolume{76},
\bfpage{023503}
(\byear{2007}).
\arxivurl{arXiv:0701380}.
doi:\doiurl{10.1103/PhysRevD.76.023503}
\end{barticle}
\endbibitem

\bibitem[\protect\citeauthoryear{{Hansen} et~al.}{2004}]{2004MNRAS.354..641H}
\begin{barticle}
\bauthor{\bsnm{{Hansen}}, \binits{F. K.}},
\bauthor{\bsnm{{Banday}}, \binits{A. J.}},
\bauthor{\bsnm{{G{\'o}rski}}, \binits{K. M.}}:
\bjtitle{\mnras}
\bvolume{354},
\bfpage{641}
(\byear{2004}).
\arxivurl{arXiv:0404206}.
doi:\doiurl{10.1111/j.1365-2966.2004.08229.x}
\end{barticle}
\endbibitem

\bibitem[\protect\citeauthoryear{{Hawking}}{1982}]{Hawking1982}
\begin{barticle}
\bauthor{\bsnm{{Hawking}}, \binits{S.W.}}:
\bjtitle{Physics Letters B}
\bvolume{115},
\bfpage{295}
(\byear{1982}).
doi:\doiurl{10.1016/0370-2693(82)90373-2}
\end{barticle}
\endbibitem

\bibitem[\protect\citeauthoryear{{Hazra} et~al.}{2013}]{2013JCAP...05..026H}
\begin{barticle}
\bauthor{\bsnm{{Hazra}}, \binits{D. K.}},
\bauthor{\bsnm{{Sriramkumar}}, \binits{L.}},
\bauthor{\bsnm{{Martin}}, \binits{J.}}:
\bjtitle{\jcap}
\bvolume{05},
\bfpage{026}
(\byear{2013}).
\arxivurl{arXiv:1201.0926}.
doi:\doiurl{10.1088/1475-7516/2013/05/026}
\end{barticle}
\endbibitem

\bibitem[\protect\citeauthoryear{{Huang} et~al.}{2015}]{2015JCAP...10..035H}
\begin{barticle}
\bauthor{\bsnm{{Huang}}, \binits{Q.-G.}},
\bauthor{\bsnm{{Wang}}, \binits{S.}},
\bauthor{\bsnm{{Zhao}}, \binits{W.}}:
\bjtitle{\jcap}
\bvolume{10},
\bfpage{035}
(\byear{2015}).
\arxivurl{arXiv:1509.02676}.
doi:\doiurl{10.1088/1475-7516/2015/10/035}
\end{barticle}
\endbibitem

\bibitem[\protect\citeauthoryear{{Ili\'c} et~al.}{2010}]{2010PhRvD..81j3502I}
\begin{barticle}
\bauthor{\bsnm{{Ili\'c}}, \binits{S.}},
\bauthor{\bsnm{{Kunz}}, \binits{M.}},
\bauthor{\bsnm{{Liddle}}, \binits{A. R.}},
\bauthor{\bsnm{{Frieman}}, \binits{J. A.}}:
\bjtitle{\prd}
\bvolume{81},
\bfpage{103502}
(\byear{2010}).
\arxivurl{arXiv:1002.4196}.
doi:\doiurl{10.1103/PhysRevD.81.103502}
\end{barticle}
\endbibitem

\bibitem[\protect\citeauthoryear{{Iqbal} et~al.}{2015}]{Iqbal2015}
\begin{barticle}
\bauthor{\bsnm{{Iqbal}}, \binits{A.}},
\bauthor{\bsnm{{Prasad}}, \binits{J.}},
\bauthor{\bsnm{{Souradeep}}, \binits{T.}},
\bauthor{\bsnm{{Malik}}, \binits{M.A.}}:
\bjtitle{JCAP}
\bvolume{06},
\bfpage{14}
(\byear{2015}).
\arxivurl{arXiv:1501.02647}.
doi:\doiurl{10.1088/1475-7516/2015/06/014}
\end{barticle}
\endbibitem

\bibitem[\protect\citeauthoryear{{Kinney}}{1997}]{1997PhRvD..56.2002K}
\begin{barticle}
\bauthor{\bsnm{{Kinney}}, \binits{W. H.}}:
\bjtitle{\prd}
\bvolume{56},
\bfpage{2002}
(\byear{1997}).
\arxivurl{arXiv:9702427}.
doi:\doiurl{10.1103/PhysRevD.56.2002}
\end{barticle}
\endbibitem

\bibitem[\protect\citeauthoryear{{Lewis} and
  {Bridle}}{2002}]{2002PhRvD..66j3511L}
\begin{barticle}
\bauthor{\bsnm{{Lewis}}, \binits{A.}},
\bauthor{\bsnm{{Bridle}}, \binits{S.}}:
\bjtitle{\prd}
\bvolume{66}(\bissue{10}),
\bfpage{103511}
(\byear{2002}).
\arxivurl{arXiv:0205436}.
doi:\doiurl{10.1103/PhysRevD.66.103511}
\end{barticle}
\endbibitem

\bibitem[\protect\citeauthoryear{{Lewis} and
  {Challinor}}{2002}]{2002PhRvD..66b3531L}
\begin{barticle}
\bauthor{\bsnm{{Lewis}}, \binits{A.}},
\bauthor{\bsnm{{Challinor}}, \binits{A.}}:
\bjtitle{\prd}
\bvolume{66}(\bissue{2}),
\bfpage{023531}
(\byear{2002}).
\arxivurl{arXiv:0203507}.
doi:\doiurl{10.1103/PhysRevD.66.023531}
\end{barticle}
\endbibitem

\bibitem[\protect\citeauthoryear{{Lewis} et~al.}{2000}]{2000ApJ...538..473L}
\begin{barticle}
\bauthor{\bsnm{{Lewis}}, \binits{A.}},
\bauthor{\bsnm{{Challinor}}, \binits{A.}},
\bauthor{\bsnm{{Lasenby}}, \binits{A.}}:
\bjtitle{\apj}
\bvolume{538},
\bfpage{473}
(\byear{2000}).
\arxivurl{arXiv:9911177}.
doi:\doiurl{10.1086/309179}
\end{barticle}
\endbibitem

\bibitem[\protect\citeauthoryear{{Liddle} and
  {Lyth}}{1993}]{1993PhR...231....1L}
\begin{barticle}
\bauthor{\bsnm{{Liddle}}, \binits{A.R.}},
\bauthor{\bsnm{{Lyth}}, \binits{D.H.}}:
\bjtitle{\physrep}
\bvolume{231},
\bfpage{1}
(\byear{1993}).
\arxivurl{arXiv:9303019}.
doi:\doiurl{10.1016/0370-1573(93)90114-S}
\end{barticle}
\endbibitem

\bibitem[\protect\citeauthoryear{{Liddle} and {Lyth}}{2000}]{Liddle2000}
\begin{bbook}
\bauthor{\bsnm{{Liddle}}, \binits{A. R.}},
\bauthor{\bsnm{{Lyth}}, \binits{D. H.}}:
\bbtitle{{Cosmological Inflation and Large-scale Structure}}.
\bpublisher{Cambridge University Press}
(\byear{2000})
\end{bbook}
\endbibitem

\bibitem[\protect\citeauthoryear{{Liddle} et~al.}{1994}]{Liddle1994}
\begin{barticle}
\bauthor{\bsnm{{Liddle}}, \binits{A. R.}},
\bauthor{\bsnm{{Parsons}}, \binits{P.}},
\bauthor{\bsnm{{Barrow}}, \binits{J. D.}}:
\bjtitle{\prd}
\bvolume{50},
\bfpage{7222}
(\byear{1994}).
\arxivurl{arXiv:9408015}.
doi:\doiurl{10.1103/PhysRevD.50.7222}
\end{barticle}
\endbibitem

\bibitem[\protect\citeauthoryear{{Lidsey} et~al.}{2004}]{2004PhRvD..70f3521L}
\begin{barticle}
\bauthor{\bsnm{{Lidsey}}, \binits{J.E.}},
\bauthor{\bsnm{{Mulryne}}, \binits{D.J.}},
\bauthor{\bsnm{{Nunes}}, \binits{N.J.}},
\bauthor{\bsnm{{Tavakol}}, \binits{R.}}:
\bjtitle{\prd}
\bvolume{70},
\bfpage{063521}
(\byear{2004}).
\arxivurl{arXiv:0406042}.
doi:\doiurl{10.1103/PhysRevD.70.063521}
\end{barticle}
\endbibitem

\bibitem[\protect\citeauthoryear{{Liu} et~al.}{2010}]{Liu2010}
\begin{barticle}
\bauthor{\bsnm{{Liu}}, \binits{Z.-G.}},
\bauthor{\bsnm{{Guo}}, \binits{Z.-K.}},
\bauthor{\bsnm{{Piao}}, \binits{Y.-S.}}:
\bjtitle{\phylB}
\bvolume{697},
\bfpage{407}
(\byear{2010}).
\arxivurl{arXiv:1012.0673}.
doi:\doiurl{10.1016/j.physletb.2010.12.055}
\end{barticle}
\endbibitem

\bibitem[\protect\citeauthoryear{{Liu} et~al.}{2014}]{2014EPJC...74.3006L}
\begin{barticle}
\bauthor{\bsnm{{Liu}}, \binits{Z.-G.}},
\bauthor{\bsnm{{Guo}}, \binits{q.K.} \bsuffix{Z}},
\bauthor{},
\bauthor{\bsnm{{Piao}}, \binits{Y.-S.}}:
\bjtitle{\epjc}
\bvolume{74},
\bfpage{3006}
(\byear{2014}).
\arxivurl{arXiv:1311.1599}.
doi:\doiurl{10.1140/epjc/s10052-014-3006-0}
\end{barticle}
\endbibitem

\bibitem[\protect\citeauthoryear{{Ludwick}}{2015}]{2015PhRvD..92f3019L}
\begin{barticle}
\bauthor{\bsnm{{Ludwick}}, \binits{K.J.}}:
\bjtitle{\prd}
\bvolume{92},
\bfpage{063019}
(\byear{2015}).
\arxivurl{arXiv:1507.06492}.
doi:\doiurl{10.1103/PhysRevD.92.063019}
\end{barticle}
\endbibitem

\bibitem[\protect\citeauthoryear{{Mortonson}
  et~al.}{2009}]{2009PhRvD..79j3519M}
\begin{barticle}
\bauthor{\bsnm{{Mortonson}}, \binits{M. J.}},
\bauthor{\bsnm{{Dvorkin}}, \binits{C.}},
\bauthor{\bsnm{{Peiris}}, \binits{H. V.}},
\bauthor{\bsnm{{Hu}}, \binits{W.}}:
\bjtitle{\prd}
\bvolume{79},
\bfpage{103519}
(\byear{2009}).
\arxivurl{arXiv:0903.4920}.
doi:\doiurl{10.1103/PhysRevD.79.103519}
\end{barticle}
\endbibitem

\bibitem[\protect\citeauthoryear{{Nojiri} and
  {Odintsov}}{2003}]{2003PhLB..562..147N}
\begin{barticle}
\bauthor{\bsnm{{Nojiri}}, \binits{S.}},
\bauthor{\bsnm{{Odintsov}}, \binits{S. D.}}:
\bjtitle{\phylB}
\bvolume{562},
\bfpage{147}
(\byear{2003}).
\arxivurl{arXiv:0303117}.
doi:\doiurl{10.1016/S0370-2693(03)00594-X}
\end{barticle}
\endbibitem

\bibitem[\protect\citeauthoryear{{Nojiri} and
  {Odintsov}}{2006}]{2006GReGr..38.1285N}
\begin{barticle}
\bauthor{\bsnm{{Nojiri}}, \binits{S.}},
\bauthor{\bsnm{{Odintsov}}, \binits{S. D.}}:
\bjtitle{Gen. Rel. Grav.}
\bvolume{38},
\bfpage{1285}
(\byear{2006}).
\arxivurl{arXiv:0506212}.
doi:\doiurl{10.1007/s10714-006-0301-6}
\end{barticle}
\endbibitem

\bibitem[\protect\citeauthoryear{{Novosyadlyj}
  et~al.}{2012}]{2012PhRvD..86h3008N}
\begin{barticle}
\bauthor{\bsnm{{Novosyadlyj}}, \binits{B.}},
\bauthor{\bsnm{{Sergijenko}}, \binits{O.}},
\bauthor{\bsnm{{Durrer}}, \binits{R.}},
\bauthor{\bsnm{{Pelykh}}, \binits{V.}}:
\bjtitle{\prd}
\bvolume{86},
\bfpage{083008}
(\byear{2012}).
\arxivurl{arXiv:0712.3328}.
doi:\doiurl{10.1103/PhysRevD.86.083008}
\end{barticle}
\endbibitem

\bibitem[\protect\citeauthoryear{{Piao}}{2008}]{2008PhRvD..78b3518P}
\begin{barticle}
\bauthor{\bsnm{{Piao}}, \binits{Y.-S.}}:
\bjtitle{\prd}
\bvolume{78},
\bfpage{023518}
(\byear{2008}).
\arxivurl{arXiv:0712.3328}.
doi:\doiurl{10.1103/PhysRevD.78.023518}
\end{barticle}
\endbibitem

\bibitem[\protect\citeauthoryear{{Piao} and
  {Zhang}}{2004}]{2004PhRvD..70f3513P}
\begin{barticle}
\bauthor{\bsnm{{Piao}}, \binits{Y.-S.}},
\bauthor{\bsnm{{Zhang}}, \binits{Y.-Z.}}:
\bjtitle{\prd}
\bvolume{70},
\bfpage{063513}
(\byear{2004}).
\arxivurl{arXiv:0401231}.
doi:\doiurl{10.1103/PhysRevD.70.063513}
\end{barticle}
\endbibitem

\bibitem[\protect\citeauthoryear{{Piao} and {Zhou}}{2003}]{2003PhRvD..68h3515P}
\begin{barticle}
\bauthor{\bsnm{{Piao}}, \binits{Y.-S.}},
\bauthor{\bsnm{{Zhou}}, \binits{E.}}:
\bjtitle{\prd}
\bvolume{68},
\bfpage{083515}
(\byear{2003}).
\arxivurl{arXiv:hep-th/0308080}.
doi:\doiurl{10.1103/PhysRevD.68.083515}
\end{barticle}
\endbibitem

\bibitem[\protect\citeauthoryear{{PLANCK Collaboration}
  et~al.}{2014a}]{Ade2013}
\begin{barticle}
\bauthor{\bsnm{{PLANCK Collaboration}}},
\bauthor{\bsnm{{Ade}}, \binits{P. A. R.}}, \betal:
\bjtitle{\aap}
\bvolume{571},
\bfpage{16}
(\byear{2014}a).
\arxivurl{arXiv:1303.5076}.
doi:\doiurl{10.1051/0004-6361/201321591}
\end{barticle}
\endbibitem

\bibitem[\protect\citeauthoryear{{PLANCK Collaboration}
  et~al.}{2014b}]{2014A&A...571A..22P}
\begin{barticle}
\bauthor{\bsnm{{PLANCK Collaboration}}},
\bauthor{\bsnm{{Ade}}, \binits{P. A. R.}},
\bauthor{\bparticle{et} \bsnm{al.}}:
\bjtitle{\aap}
\bvolume{571},
\bfpage{22}
(\byear{2014}b).
\arxivurl{arXiv:1303.5082}.
doi:\doiurl{10.1051/0004-6361/201321569}
\end{barticle}
\endbibitem

\bibitem[\protect\citeauthoryear{{PLANCK Collaboration}
  et~al.}{2014c}]{Ade2013b}
\begin{barticle}
\bauthor{\bsnm{{PLANCK Collaboration}}},
\bauthor{\bsnm{{Ade}}, \binits{P. A. R.}}, \betal:
\bjtitle{\aap}
\bvolume{571},
\bfpage{23}
(\byear{2014}c).
\arxivurl{arXiv:1303.5083}.
doi:\doiurl{10.1051/0004-6361/201321534}
\end{barticle}
\endbibitem

\bibitem[\protect\citeauthoryear{{PLANCK Collaboration}
  et~al.}{2016a}]{Ade2015}
\begin{barticle}
\bauthor{\bsnm{{PLANCK Collaboration}}},
\bauthor{\bsnm{{Ade}}, \binits{P. A. R.}}, \betal:
\bjtitle{\aap}
\bvolume{594},
\bfpage{13}
(\byear{2016}a).
\arxivurl{arXiv:1502.01589}.
doi:\doiurl{10.1051/0004-6361/201525898}
\end{barticle}
\endbibitem

\bibitem[\protect\citeauthoryear{{PLANCK Collaboration}
  et~al.}{2016b}]{Ade2015c}
\begin{barticle}
\bauthor{\bsnm{{PLANCK Collaboration}}},
\bauthor{\bsnm{{Ade}}, \binits{P. A. R.}}, \betal:
\bjtitle{\aap}
\bvolume{594},
\bfpage{16}
(\byear{2016}b).
\arxivurl{arXiv:1506.07135}.
doi:\doiurl{10.1051/0004-6361/201526681}
\end{barticle}
\endbibitem

\bibitem[\protect\citeauthoryear{{PLANCK Collaboration}
  et~al.}{2016c}]{Ade2015b}
\begin{barticle}
\bauthor{\bsnm{{PLANCK Collaboration}}},
\bauthor{\bsnm{{Ade}}, \binits{P. A. R.}}, \betal:
\bjtitle{\aap}
\bvolume{594},
\bfpage{20}
(\byear{2016}c).
\arxivurl{arXiv:1502.02114}
\end{barticle}
\endbibitem

\bibitem[\protect\citeauthoryear{{PLANCK Collaboration}
  et~al.}{2018}]{2018arXiv180706209P}
\begin{botherref}
\oauthor{\bsnm{{PLANCK Collaboration}}},
\oauthor{\bsnm{{Ade}}, \binits{P. A. R.}}, et al.:
{Planck 2018 results. VI. Cosmological parameters}
(2018).
\arxivurl{arXiv:1807.06209}
\end{botherref}
\endbibitem

\bibitem[\protect\citeauthoryear{{Pollock}}{1988}]{1988PhLB..215..635P}
\begin{barticle}
\bauthor{\bsnm{{Pollock}}, \binits{M. D.}}:
\bjtitle{\phylB}
\bvolume{215},
\bfpage{635}
(\byear{1988}).
doi:\doiurl{10.1016/0370-2693(88)90034-2}
\end{barticle}
\endbibitem

\bibitem[\protect\citeauthoryear{{Qureshi} et~al.}{2017}]{2017JCAP...04..013Q}
\begin{barticle}
\bauthor{\bsnm{{Qureshi}}, \binits{M. H.}},
\bauthor{\bsnm{{Iqbal}}, \binits{A.}},
\bauthor{\bsnm{{Malik}}, \binits{M. A.}},
\bauthor{\bsnm{{Souradeep}}, \binits{T.}}:
\bjtitle{\jcap}
\bvolume{04},
\bfpage{013}
(\byear{2017}).
\arxivurl{arXiv:1610.05776}.
doi:\doiurl{10.1088/1475-7516/2017/04/013}
\end{barticle}
\endbibitem

\bibitem[\protect\citeauthoryear{{Richarte} and
  {Kremer}}{2017}]{2017EPJC...77...51R}
\begin{barticle}
\bauthor{\bsnm{{Richarte}}, \binits{M. G.}},
\bauthor{\bsnm{{Kremer}}, \binits{G. M.}}:
\bjtitle{\epjc}
\bvolume{77},
\bfpage{51}
(\byear{2017}).
\arxivurl{arXiv:1612.03822}.
doi:\doiurl{10.1140/epjc/s10052-017-4629-8}
\end{barticle}
\endbibitem

\bibitem[\protect\citeauthoryear{{Sahni} and
  {Shtanov}}{2003}]{2003JCAP...11..014S}
\begin{barticle}
\bauthor{\bsnm{{Sahni}}, \binits{V.}},
\bauthor{\bsnm{{Shtanov}}, \binits{Y.}}:
\bjtitle{\jcap}
\bvolume{11},
\bfpage{014}
(\byear{2003}).
\arxivurl{arXiv:0202346}.
doi:\doiurl{10.1088/1475-7516/2003/11/014}
\end{barticle}
\endbibitem

\bibitem[\protect\citeauthoryear{{Sami} and
  {Toporensky}}{2004}]{2004MPLA...19.1509S}
\begin{barticle}
\bauthor{\bsnm{{Sami}}, \binits{M.}},
\bauthor{\bsnm{{Toporensky}}, \binits{A.}}:
\bjtitle{\phylA}
\bvolume{19},
\bfpage{1509}
(\byear{2004}).
\arxivurl{arXiv:0312009}.
doi:\doiurl{10.1142/S0217732304013921}
\end{barticle}
\endbibitem

\bibitem[\protect\citeauthoryear{{Seljak} and
  {Zaldarriaga}}{1996}]{1996ApJ...469..437S}
\begin{barticle}
\bauthor{\bsnm{{Seljak}}, \binits{U.}},
\bauthor{\bsnm{{Zaldarriaga}}, \binits{M.}}:
\bjtitle{\apj}
\bvolume{469},
\bfpage{437}
(\byear{1996}).
\arxivurl{arXiv:9603033}.
doi:\doiurl{10.1086/177793}
\end{barticle}
\endbibitem

\bibitem[\protect\citeauthoryear{{Singh} et~al.}{2003}]{2003PhRvD..68b3522S}
\begin{barticle}
\bauthor{\bsnm{{Singh}}, \binits{P.}},
\bauthor{\bsnm{{Sami}}, \binits{M.}},
\bauthor{\bsnm{{Dadhich}}, \binits{N.}}:
\bjtitle{\prd}
\bvolume{68},
\bfpage{023522}
(\byear{2003}).
\arxivurl{arXiv:hep-th/0305110}.
doi:\doiurl{10.1103/PhysRevD.68.023522}
\end{barticle}
\endbibitem

\bibitem[\protect\citeauthoryear{{Smoot} et~al.}{1992}]{Smoot1992}
\begin{barticle}
\bauthor{\bsnm{{Smoot}}, \binits{G.F.}}, \betal:
\bjtitle{\apjl}
\bvolume{396},
\bfpage{1}
(\byear{1992}).
doi:\doiurl{10.1086/186504}
\end{barticle}
\endbibitem

\bibitem[\protect\citeauthoryear{Weinberg}{1972}]{Weinberg1972}
\begin{bbook}
\bauthor{\bsnm{Weinberg}, \binits{S.}}:
\bbtitle{{Gravitation and Cosmology: Principles and Applications of the General
  Theory of Relativity}}.
\bpublisher{WILEY}
(\byear{1972})
\end{bbook}
\endbibitem

\bibitem[\protect\citeauthoryear{{WMAP Collaboration}
  et~al.}{2013}]{Hinshaw2013}
\begin{barticle}
\bauthor{\bsnm{{WMAP Collaboration}}},
\bauthor{\bsnm{{Hinshaw}}, \binits{G.}}, \betal:
\bjtitle{\apjl}
\bvolume{208},
\bfpage{19}
(\byear{2013}).
\arxivurl{arXiv:1212.5226}.
doi:\doiurl{10.1088/0067-0049/208/2/19}
\end{barticle}
\endbibitem

\bibitem[\protect\citeauthoryear{{Wu} and {Yu}}{2006}]{2006JCAP...05..008W}
\begin{barticle}
\bauthor{\bsnm{{Wu}}, \binits{P.}},
\bauthor{\bsnm{{Yu}}, \binits{H.}}:
\bjtitle{\jcap}
\bvolume{05},
\bfpage{08}
(\byear{2006}).
\arxivurl{arXiv:gr-qc/0604117}.
doi:\doiurl{10.1088/1475-7516/2006/05/008}
\end{barticle}
\endbibitem

\end{thebibliography}

\end{document}